# A Bleeding Digital Heart: Identifying Residual Data Generation from Smartphone Applications Interacting with Medical Devices

George Grispos
School of Interdisciplinary Informatics
College of Information Science and Technology
University of Nebraska at Omaha
ggrispos@unomaha.edu

William Bradley Glisson
Cyber Forensics Intelligence Center, Department of Computer Science
Sam Houston State University
glisson@shsu.edu

Peter Cooper
Cyber Forensics Intelligence Center, Department of Computer Science
Sam Houston State University
csc_pac@shsu.edu

## Abstract

*The integration of medical devices in everyday life prompts the idea that these devices will increasingly have evidential value in civil and criminal proceedings. However, the investigation of these devices presents new challenges for the digital forensics community. Previous research has shown that mobile devices provide investigators with a wealth of information. Hence, mobile devices that are used within medical environments potentially provide an avenue for investigating and analyzing digital evidence from such devices.*

*The research contribution of this paper is twofold. First, it provides an empirical analysis of the viability of using information from smartphone applications developed to complement a medical device, as digital evidence. Second, it includes documentation on the artifacts that are potentially useful in a digital forensics investigation of smartphone applications that interact with medical devices.*

## 1. Introduction

The integration of information technology into medical settings introduces a variety of opportunities and challenges from security and investigation perspectives. Medical devices such as ventilators, pacemakers and infusion pumps, which initially operated as standalone devices, now integrate network technology that permits communication with a variety of connected healthcare systems [1]. This integration has resulted in several benefits including improved patient safety, reduced workloads for medical professionals, and the increased generation of medical data [2, 3].

According to Qualcomm Life [4, 5], next-generation medical devices will produce and collect vast amounts of medical and patient information. For example, a typical ventilator is expected to generate almost 305 data parameters per second, which includes patient information, therapy details, and device operation metadata [4, 5]. Based on these predictions, a Stanford Medicine report goes on to estimate that the medical industry will generate 2,314 exabytes of data by 2020 [6]. From a digital forensics' perspective, medical device data could provide a forensic investigator with a wealth of potential evidence.

Previous research indicates that digital evidence is becoming increasingly important in a variety of legal situations [7-9]. Hence, it is only a matter of time before medical device data becomes prominent in civil and criminal cases. Support for these predictions is already evident in criminal cases in the United States. For example, in Middletown Ohio, law enforcement investigators issued a search warrant to collect information from an individual's pacemaker [10]. This information was then used to charge the individual with a variety of crimes, including arson. Similarly, a court in Pennsylvania admitted evidence from a fitness watch device to support charges of false reports to law enforcement, false alarms to public safety, and tampering with evidence [11]. However, the digital evidence was only analyzed by investigators after the defendant provided their username and password.

Hence, defendants who do not provide device authentication information introduce additional barriers for digital forensic investigators. Consequently, researchers have speculated that the collection and analysis, of residual data from medical devices, are unlikely to be straightforward [2]. The underlying problem is the absence of digital forensic consideration in the design of medical products, introducing uncertainty as to the viability of conventional approaches to forensic evidence acquisition in medical environments [2].

Traditionally, when a storage device is involved in a crime, the equipment is seized to obtain an accurate copy of the data stored on the device for later analysis [12]. Typically, devices connect to an investigator's computer via a write-blocker. An investigator then creates a byte-for-byte copy (an image) of the entire storage device using commercial software tools, such as FTK or Encase, or an open source alternative like 'dd'. During the analysis phase of an investigation, an investigator appraises the significance of extracted information artifacts. The investigator then develops a narrative based on the extracted residual data and constructs a timeline using the data to explain how a crime was potentially committed. Where appropriate, it may also be necessary to associate particular artifacts with one or more user IDs. Evidence recovered from traditional storage media (e.g., documents, images, multimedia files, and metadata) is known to assist with this type of analysis. However, it is not clear that the residual data recovered from medical devices and their associated applications will support similar analysis activities. The overall lack of clarity is due to the variety of medical devices available in the market in conjunction with the deficiency of knowledge about the 'forensic-ability' (i.e., the existence and availability of timestamp information and relevant metadata) of medical device evidence.

One medical device that has received particular attention in the past twelve months is the Kardia Mobile [13-15]. Kardia is a hand-held single-lead Electrocardiography (ECG) device that utilizes a smartphone to detect and monitor Atrial Fibrillation (AF) [15]. From a safety perspective, Kardia and its associated smartphone application are approved for use by both the United States Food And Drug Administration and the European Union [16]. The device is intended to assist medical professionals in the treatment of patients with known or suspected heart conditions, such as AF [15, 16].

The increased usage of residual data in court cases in conjunction with recent legal activities supports the idea that approved medical devices, such as the Kardia, will escalate in importance in legal contexts. It is also realistic to speculate that as network connectivity and digital functionality increase in medical devices, that these devices will increasingly become cybersecurity targets. This line of thought prompts the hypothesis *that a medical device smartphone application can provide forensically relevant data.* To address the hypothesis, the following questions were proposed:
- Can metadata, concerning a patient and their use of the medical device be recovered from a smartphone application?
- Can medical information be recovered from an accompanying smartphone application?

The research contribution of this paper is twofold. First, it provides an empirical analysis of the viability of using information from smartphone applications developed to complement a medical device, as evidence in legal cases. Second, it provides documentation on the artifacts that are potentially useful in a digital forensics investigation of smartphone applications that interact with medical devices. This paper is structured as follows. Section two discusses relevant previous research related to medical devices and smartphone forensics. Section three presents the experiment design implemented in this research. Section four reports the experimental findings, and a discussion of the results. Finally, Section five draws conclusions from the work conducted and presents ideas for future research.

## 2. Related Work

Researchers are continuously demonstrating that medical devices are at risk from a security perspective [17-21]. Malasri and Wang [17] examined the insecurity of implantable medical devices, such as pacemakers. These researchers argue that implantable medical devices are susceptible to a variety of attacks including eavesdropping, patient tracking, and spoofing. They concluded that a malicious attacker could, in theory, cause direct physical harm to an individual by sending malicious commands to compromise the security of such devices. Glisson et al. [18] examined the viability of compromising a medical mannequin, in a live hospital environment. In addition to demonstrating two types of brute force attacks, these researchers also explained how a medical mannequin is vulnerable to a denial of service attack. Li et al. [19] focused their research on a diabetes therapy device and reported that some medical devices transmit patient and device information in plaintext. After analyzing transmitted information, they reported that they were able to recover device passwords and network data packets containing dosage information.

Growing security concerns with medical devices have prompted the application of increased regulatory pressure on healthcare organizations to require forensic investigation capabilities [22]. Digital forensics is concerned with the collection, analysis, and admissibility of collected data in a legal context [23]. Forensic investigations attempt to establish answers to five key questions, known as the five W's: what, why, who, when, and where [23]. From a medical device perspective, researchers have raised concerns regarding the forensic investigation of such devices [1, 2, 18, 24, 25].

Grispos et al. [2] argue that in order to answer the five W's, investigators rely on the residual data from devices and systems involved in a crime. However, they speculate that such data might not always be available within medical devices, and as a result, investigators might not be able to identify suspects or facts within an investigation. In line with these thoughts, Glisson et al. [18] indicated that the digital forensics community required further research to identify medical device residual data, which could be used in a forensic investigation.

Cusack and Kyaw [1] proposed an alternative network-centric solution based on the integration of forensic capabilities during the development of medical devices and systems. They proposed a forensically-ready hospital wireless network architecture but do not attempt to address residual data generated by medical devices. Ellouze et al. [24] focused on the type of evidence that is needed in medical device investigations, including data collected from sensor nodes, mobile devices, wireless medical devices, databases, and third-party applications. In later work, Ellouze et al. [25] proposed a theoretical investigation framework to help guide a post-mortem analysis of implantable medical devices. Part of this solution involves developing a set of techniques to assist with the secure storage of evidence logs on mobile devices, which can be used to track sensitive medical events.

When mobile devices are used within medical environments, they potentially provide an avenue for investigating and analyzing digital evidence. Mobile device forensics is defined as "the science of recovering digital evidence from a mobile device under forensically sound conditions using acceptable methods" [26]. From a digital forensics perspective, a mobile device can be considered a treasure trove of forensic evidence [27-29]. A study by Glisson et al. [27] involving the analysis of 49 low-end devices acquired from second-hand markets, recovered more than 11,000 artifacts. However, the advancement of smartphones has resulted in the generation of richer artifacts such as web-browsing activities, third-party application data and GPS coordinates [28]. As a result, several researchers have focused their efforts on understanding the residual artifacts generated on smartphone devices.

Grispos et al. [30, 31] and Martini et al. [32] analyzed mobile cloud storage applications on Android and iOS devices. This analysis included the identification of data and metadata, along with their locations on the mobile device. Al Mutawa et al. [33] examined social networking applications, such as Facebook, Twitter, and MySpace, on Android, iOS, and Blackberry devices. While minimal artifacts could be recovered from Blackberry devices, the researchers observed that evidentiary artifacts concerning these applications could be recovered from Android and iOS-based devices. Anglano [34] focused on examining the artifacts generated by users who use the communication application, WhatsApp. The findings from this research show that artifacts could be recovered from databases, which can be used to reconstruct the list of contacts and the chronology of the messages that have been exchanged between users. Badar and Baggili [35] examined the logical backup generated from an Apple iPhone when connected to iTunes, with the goal of using this backup as a source of forensic artifacts. Using this approach, Badar and Baggili [35] recovered a variety of iPhone-related artifacts including the device's address book, incoming and outgoing call logs, text messages, email account information, and internet browser history. Levinson et al. [28] examined partitions on an iPhone device with the aim of identifying residual data from built-in applications. From an artifact perspective, Levinson et al. [28] recovered user account information, timestamps, geolocation information, contacts and multimedia files from the device.

In addition to the above applications and operating systems, researchers have focused on the examination of Mobile Health (mHealth) and well-being applications. Plachkinova et al. [36] focused on the security and privacy of 38 Android and iOS mHealth applications with the objective of developing a taxonomy based on security and privacy issues identified from the applications. Kharrazi et al. [37] examined nineteen mHealth applications on iOS, Blackberry, and Android devices and reported that seven of the evaluated applications did not implement security features such as password protection and user authentication. From a forensics perspective, Azfar et al. [38] examined 40 Android mHealth applications and proposed a forensic taxonomy that comprises databases, user credentials, personal details of users, user activities, user location, activity timestamps, and images. While previous research has examined residual, security and privacy issues in medical devices and mHealth applications, minimal research examines the generation of smartphone application residual data as a consequence of interacting with a medical device.

## 3. Experiment Design

In order to answer the hypothesis and associated research questions identified in the introduction, a controlled experiment was undertaken [39]. This controlled experiment consisted of five stages. The five stages included: 1) preparing a smartphone device for use in the experiment; 2) installing the Kardia application onto a smartphone device and setting up a test account for use in the experiment; 3) using the Kardia application and the Kardia Mobile ECG device; 4) processing the smartphone device using MicroSystemation (MSAB) XRY; and 5) using forensic tools to extract files and artifacts from the extraction dumps.

The mobile device forensic toolkit used in this experiment was MSAB's XRY version 7.7 and its associated analysis tool, XAMN version 3.2 [40]. Files and artifacts associated with the Kardia application were then extracted from the memory dumps using XAMN and analyzed further using AccessData Forensic Toolkit (FTK) version 6.3 [41]. These tools were chosen based on practicality and availability to the authors.

Two smartphone devices were included in this experiment: a Samsung Galaxy S4 and an Apple iPhone SE (hereafter referred to as the devices). Table 1 - Smartphone Devices, presents an overview of these devices, their features and storage capabilities. These devices were selected based on their compatibility with the XRY forensic toolkit that was used to extract a memory dump of the device's internal memory. In addition, the Android and iOS operating systems executed on the devices represent the two most popular smartphone operating systems, at the time of the research [42]. A number of smartphone devices fulfill these criteria and could have been used in the research. The decision to use these specific devices was based on availability.

| Feature | Galaxy S4 | iPhone SE |
|---|---|---|
| Model Number | SGH-i337 | A1662 |
| Operating System | Android v. 5.0 (Lollipop) | iOS v. 11.3.1 |
| Storage Capacity | 16 GB | 32 GB |

**Table 1: Smartphone Devices**

The smartphone application used in this experiment was Kardia version 5.1.2, for both Android and iOS operating systems. The following steps were undertaken to prepare the devices and applications for the experiment. These steps were executed for both the Android and iOS smartphone devices.

1. The device was 'hard reset' to remove any previous data. This involved restoring the factory settings on the device. A desktop computer was used to create, depending on the device, a Google or Apple account. The device was then powered-on, and the account was used to complete the initial device setup. All default device setup options were selected during this initial process.

2. The device was then connected to a wireless network, which was used to gain access to the Internet. The Kardia application was downloaded and installed using the device's respective application store (Google Play and the Apple App Store). The default installation and security parameters were used during the installation of the application. The Kardia application was executed, and a new profile was created using an email address created for the purpose of the experiment and a common password. The profile information recorded included: first name, last name, date of birth, height, gender, and smoker status. Test information was used for the purpose of completing the profile. After the initial setup, the default settings were used on the Kardia application. This included settings for restricting an ECG recording to thirty seconds, permitting voice notes and enabling a thirty-day premium service trial.

3. After the Kardia application setup was completed, the application was used to record an initial measurement using the Kardia Mobile device. This initial measurement is required before the Kardia Mobile and application can be used by a patient. The measurement is recorded and then submitted to a medical professional by the application, on behalf of the patient. After a notification was received that the initial recording was successful, the Kardia Mobile and the application were then used twice a day for five days. The date and time of each recording, along with the 'result' provided on the smartphone interface by the Kardia application were documented for future reference. In addition to recording ECG measurements, the smartphone application interface was also used to manually record blood pressure recordings and weight information. Test information was used for the purpose of these measurements, which were documented for future analysis.

4. Upon completion of the five-day period, the smartphone device was processed using the XRY toolkit to create a memory dump of its internal memory. In the case of the Galaxy S4, the device was processed to create a 'Physical Acquisition' of the internal memory. This extraction process first involved activating the 'USB Debugging' option on the smartphone device. A step-by-step wizard provided instructions on how to prepare the device for the extraction, and a number of forensic programs were then loaded onto the smartphone. The Galaxy S4's internal memory was then read, and a memory dump was saved to a folder on the desktop of a forensic workstation. The entire process took approximately forty minutes.

The extraction process for the Apple SE differed from that of the Galaxy S4. In the case of the Apple SE, only a 'Full Read Logical' extraction was possible using the XRY forensic toolkit. Nonetheless, this type of extraction can still be used to recover the Kardia application folder located within the User partition on the Apple Device. A step-by-step wizard provided instructions on how to prepare the device for the logical acquisition. The resulting extraction was then saved on the desktop of a forensic workstation. The entire process took approximately fifteen minutes.

5. The subsequent forensic extractions that were created using the XRY were then loaded in XAMN, where the Android and iOS file systems were reconstructed. The analysis techniques used to locate files and artifacts related to the Kardia application included: string searching, text filtering, and browsing the respective file systems. Files and artifacts were also extracted from the memory dumps using XMAN and analyzed further using FTK.

The scope of this research is restricted in the following ways. The experiment was conducted in the United States (US) using devices that contain network software for carrier providers in the US. The experiment was executed only once on each device. The experiment focused on a specific version of the Android and iOS operating systems, a specific version of the Kardia application, and specific versions of XRY and FTK. Due to tool limitation, a Logical Extraction was the sole extraction method for the iOS device, while a Physical Extraction was the extraction method for the Android device. It should also be noted that the primary researcher was both a participant and a researcher in the experiment.

## 4. Results and Discussion

An analysis of the Android and iOS memory dumps revealed that a variety of artifacts related to the Kardia Mobile could be recovered from its smartphone application. These artifacts can be broadly categorized as medical-type data, patient information, and user-generated metadata.

### 4.1. Kardia Application on Android

Within an Android device, user-related files and metadata are stored under the location /data/data [43]. Kardia Mobile artifacts were recovered from an application folder called com.alivecor.aliveecg that was stored under the /data/data location. Medical-type artifacts can be recovered from various locations in the application folder. Within the high-level folder is a subfolder called Files. This contains various subfolders of potential interest to a forensic investigation. ECG files generated by the Kardia Mobile device can be recovered from the location: com.alivecor.aliveecg/files/ecgs. Two ECG files are generated for each reading taken with the Kardia, and both files contain an .atc file extension. Figure 1 presents an extract of the file header information for these files using FTK. This file header information contains the date and time the reading was taken. It also contains the ECG reading's identifier number (3db73498-32a0-4293-b5f0-7616162c55d8), the smartphone device used to take the recording, and the version of Kardia Mobile.

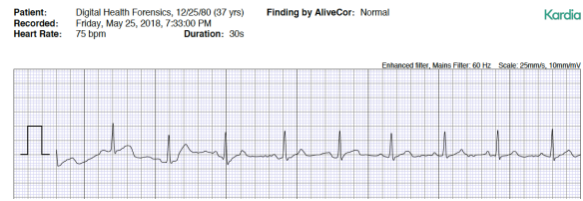

**Figure 1: Android .ATC File Header**

An analysis of the subfolder called temp under the folder com.alivecor.aliveecg/files/ revealed PDF documents. These PDF documents contain any ECG recordings that are referred to a medical professional for further analysis. Figure 2 presents an extract of this PDF document, which includes patient information, the patient's date of birth, when the recording was undertaken, and the heart rate documented as beats per minute.

**Figure 2: PDF of ECG Recording**

In addition to the ECG recordings, audio files related to the recordings can be found in the location: com.alivecor.aliveecg/files/. These files are in two formats: Advanced Audio Coding (.aac) and MPEG 4 Audio (.m4a). The files are thirty seconds long and record audio sounds during the ECG recording. The audio file name can be used to look up the corresponding ECG recording using the UUID value in the database discussed below.

Patient information and metadata related to the Kardia can be recovered from a database called ECG.db. This database contains thirteen tables, documenting various metadata and can be recovered from the location: `com.alivecor.aliveecg/databases`. The table `bp_records` contains metadata related to blood pressure recordings. This includes timestamp information, timestamp offsets, if the record is deleted or not, systolic and diastolic readings, the heart rate at the time of the recording and the source of the blood pressure recording. It must be noted that the timestamp information was not in the local time but stored as an epoch timestamp according to Greenwich Mean Time (GMT). As a result, the timestamp offset (-18000 seconds) must be applied to the documented timestamp to obtain the actual day and time associated with the recording.

Within the `ECG` table, metadata related to the patient and their ECG recordings can be recovered. This metadata includes the patient's first and last name, as well as their date of birth (stored as a GMT epoch timestamp). The table also stores information related to ECG readings taken using the Kardia Mobile device. Metadata information related to these ECG readings is summarized in Table 2 below.

| Field Name(s) | Description |
|---|---|
| uuid and server_id | Recording ID numbers |
| data Recorded | Date and time that the ECG recording was undertaken |
| duration | Duration in milliseconds of each ECG recording |
| heart_rate | Patient's heart rate in beats per minute |
| inverted | The position of the Kardia Mobile at the time of the reading: upside down (1) or correct side up (0) |
| height | Patient's height in centimeters |
| weight | Patient's weight in kilograms |
| gender | Patients gender: female (0) or male (1) |
| smoker | As entered by the patient, non-smoker (0) or smoker (1) |
| has_audio_file | If an audio file is available (1) or not available (0) |

**Table 2: Android ECG Recording Fields**

In addition to recovering ECG recordings, a third table called `Orders` can be used to identify which ECG recordings are submitted to a medical professional for further analysis. Metadata related to this table includes the ID number of the ECG recording submitted to the medical professional, the result sent to the patient, timestamp information describing when the analysis was requested, along with timestamp information when the analysis was received by the Kardia smartphone application.

The fourth table within the `ECG.db` database that contains information of potential forensic interest is called `Weight_records`. The purpose of this table is to record the patient's weight information, as entered through the Kardia application. Metadata, which can be recovered from this table includes timestamp information in epoch format (the offset must be applied to obtain the local timestamp), the weight in kilograms, the patient's height in centimeters and the source of the information. An interesting observation is that the weight and height information recovered from the database tables was in kilograms and centimeters. However, at the time of the experiment, this information was input into the smartphone interface in pounds and feet, respectively. This would suggest that the smartphone application is converting this information from one format to another, before storing it in the database table.

The Kardia application also stores patient metadata within Extensible Markup Language (XML) files. These XML files are stored in a subfolder called `shared_prefs` within the main folder located at `com.alivecor.aliveecg/`. Within `shared_prefs` the following files and information can be recovered:

- `com.alivecor.aliveecg_preferences.xml` – the email account used with the Kardia service, timestamp information related to the last blood pressure recording, last weight recording and last heart rate recording.

- `com.google.android.gms.measurement.prefs.xml` – epoch timestamp information about when the Kardia application was first used.

- `userprofile.xml` – the patient's first and last name, the patient's date of birth, the patient's weight in kilograms, the patient's email address, the patient's country location, and the patient's smoker status.

### 4.2. Kardia Application on iOS

An analysis of the iOS memory dumps revealed that Kardia device artifacts are stored under the location `/private/var/mobile/containers/data/application` in a folder called `com.alivecor.professional.aliveecg`. The primary location contains two subfolders called `Documents` and `Library`, which contain medical-type artifacts and patient metadata. Unless stated, all timestamp information recovered from the iOS application was in local Mac Absolute Time format [44]. The format of this timestamp is the number of seconds since midnight January 1, 2001, GMT [44].

Within the `Documents` subfolder, are files containing the audio sounds detected during the ECG recording. These files are in the MPEG 4 Audio (.m4a) format. The file name of the audio file corresponds to the `ZUUID` metadata, which can be recovered from the `AliveECGDB.sqlite` database. This can be used to match the audio file with the corresponding ECG reading.

A subfolder within Documents called `ecgfiles` contains the Kardia native .atc files for each ECG reading undertaken using the device. The file headers for the .atc files recovered from the Apple device are in the same format as those recovered from the Android device. The header information that can be recovered includes the data and time of the ECG recording, the ECG reading's identifier and the version of iOS running on the Apple device used with the reading.

The primary metadata location for the iOS Kardia application is an SQLite database called `AliveECGDB.sqlite`. This database can be found in the `Documents` folder and contains various tables of interest to a forensic investigation. The `ZKDMBLOODPRESSURERECORDING` table found in the database contains the metadata associated with blood pressure recordings. Metadata recovered from the table includes the systolic and diastolic values, timestamp information, notes documented at the time of the measurement and the heart rate. The `ZKDMWEIGHT` table within the `AliveECGDB.sqlite` database contains the metadata associated with weight recordings. Metadata recovered from the table includes the patient's height in centimeters, timestamp information, the patient's weight in kilogram s, and the information source or if the patient manually entered the information.

Metadata associated with any ECG reading that has been submitted to a medical professional can be recovered from a table called `ZOVERREADERORDER`. Metadata recovered from this table includes the timestamp of when the recording was submitted for further analysis, timestamp information of when the medical professional completed their analysis, and the results of this analysis. In addition, the ECG recording ID (`Z_PK`) can be used to look up more information about the submitted recording in the `ZECG` table.

The largest table within the `AliveECGDB.sqlite` database is called `ZECG`. This table contains medical information (Table 3) and patient metadata (Table 4) related to the ECG recordings undertaken using the iOS application.

| Field Name | Description |
|---|---|
| ZDATERECORDED | Timestamp that describes when the ECG reading was undertaken |
| ZHEARTRATE | The results of the ECG reading undertaken with the Kardia, stored in beats per minute |
| ZCOMMENT | Text describing the audio recording taken at the time of the ECG reading |
| ZFILENAME | The file name of the .atc file associated with the ECG reading |
| ZDATESYNCED | Timestamp that describes when the ECG reading was synchronized with a Kardia sever |
| ZDURATION_MS | Duration of the ECG recording in milliseconds |
| ZMC_ANGINA | Describes if the patient has selected 'Angina' as a medical condition in their profile (1 = YES, 0 = NO) |
| ZHAS_AUDIO_DESCRIPTION | Describes if the ECG reading includes an audio file (1 = YES, 0 = NO) |
| ZINVERTED | The position of the Kardia Mobile at the time of the reading: upside down (1) or correct side up (0) |
| Z_IS_RESTING_HEART_RATE | Describes if the ECG reading taken was 'at rest' (1 = YES, 0 = NO) |

**Table 3: iOS ECG Medical Information**

The iOS Kardia application also stores patient metadata within `Property List (plist)` files. These `plist` files are stored in a subfolder called `Preferences` within the `Library` folder. Within the `Preferences` subfolder, only one file contains forensically-relevant information called `com.alivecor.professional.aliveecg.plist`. This file contains the following information:

- the patient's first name, last name, gender and date of birth,
- the patient's height in centimeters,
- the patient's email address,
- the medical conditions selected in the profile, and
- application version and timestamp information.

| Field Name | Description |
|---|---|
| ZUUID | ID numbers assigned to each recording |
| ZMALE | Denotes if the patient is Male (1) or Female (0) |
| ZPATIENTFIRSTNAME | The patient's first name, as recorded in the patient's profile |
| ZPATIENTLASTNAME | The patient's surname name, as recorded in the patient's profile |
| ZPATIENTDOB | The patient's date of birth, as recorded in the patient's profile |
| ZHEIGHT | The patient's height, as recorded in the patient's profile |

**Table 4: iOS ECG Patient Information**

### 4.3. Analysis Summary

The results indicate that the research questions identified in the introduction, which focus on patient metadata and medical information, are acquirable through application data extraction analysis. The extracted residual data identified metadata specific to patients and their use of the medical device from the smartphone application. This metadata included the patient's first and last name, the patient's gender, the patient's height and weight, along with timestamp information regarding when the patient undertook ECG and blood pressure readings. An interesting observation from the analysis is that some of the timestamp information recovered from the artifacts were in the local time zone, while other timestamp information was stored in the Greenwich Mean Time (GMT) zone. This scenario could potentially complicate a forensic investigation, especially when an investigator is attempting to undertake a timeline analysis to determine when particular events took place.

The results from the controlled experiment have shown that it is possible to recover medical information from the Kardia Mobile application. On the Android and iOS devices, this medical information included heart rates through ECG readings, systolic and diastolic blood pressure readings, and PDF documents of ECG readings. In addition, the analysis of databases created by the Kardia application revealed that it was possible to recover medical conditions, smoker status, and audio recordings taken at the time of ECG readings.

The overall analysis of the data supports the hypothesis that medical device smartphone applications can provide relevant forensic data. This statement holds for both the Android and iOS devices analyzed during the experiment. A forensic investigator could use the medical information and patient metadata generated by the Kardia application as a potential source of digital evidence, when an investigation requires such information.

### 5. Conclusion and Future Work

Digital evidence is becoming increasingly important in a variety of legal situations. The amalgamation of medical devices into everyday life is creating a scenario where digital evidence from these devices could plausibly be introduced into civil or criminal proceedings. Hence, there is a growing interest from both industry and academia to identify information that can be recovered from these devices that will enable it to be submitted as evidence to a court of law. The results of this preliminary research demonstrate that smartphone applications that interact with medical devices provide an avenue for obtaining digital evidence from these devices. The recovery of data from these medical device applications could be used to provide a forensic investigator with information about the health and well-being of an individual involved, or the medical state of an individual at a particular time. Such evidence could be critical in criminal investigations where foul-play or medical malpractice is suspected.

Future work will examine several key areas including analysis of other medical devices, operating systems and investigating medical devices themselves for forensic evidence. Further research will investigate a greater variety of medical devices and their accompanying smartphone applications. Further research would also need to consider an examination of smartphone applications from multiple operating systems.

Future research will evaluate the results of this study on a larger scale. The research will be expanded to include the identification of storage and usage patterns along with the implementation of machine learning algorithms for the purposes of developing a user profile. The idea behind the digital profile development is to attempt to discover ways to tie individuals to profiles and profiles to devices.

An additional avenue of research will investigate the extent to which digital evidence can be recovered from a medical device itself. The purpose of this research is to investigate the relationship between data stored in a smartphone application and the data stored in a physical medical device. In theory, the extraction of both the device and associated applications should provide a more complete dataset from the physical medical device. Research also needs to consider if medical device development should integrate forensic-by-design principles in order to enhance forensic readiness and enable additional data collection for investigations involving medical devices.

## 6. Acknowledgments

Financial support for this research is provided by the Nebraska Research Initiative (NRI). Idea development is supported by the Digital Forensics and Cybersecurity Research Center at Sam Houston State University.